\documentclass[aps,prb,amsmath,
twocolumn,groupedaddress]{revtex4}
\usepackage{latexsym}
\usepackage{amsmath}
\usepackage{graphicx}
\begin{document}

\title{Spin Transport in Disordered Two-Dimensional Hopping Systems with
Rashba Spin-Orbit Interaction}
\author{U. Beckmann}
\affiliation{Institute for Theoretical Physics, 
Otto-von-Guericke-University, PF 4120, D-39016 Magdeburg, Germany}
\author{T. Damker}
\email{thomas.damker@physik.uni-magdeburg.de}
\affiliation{Institute for Theoretical Physics, 
Otto-von-Guericke-University, PF 4120, D-39016 Magdeburg, Germany}
\author{H. B\"ottger}
\affiliation{Institute for Theoretical Physics, 
Otto-von-Guericke-University, PF 4120, D-39016 Magdeburg, Germany}
\date{\today}

\begin{abstract}
The influence of Rashba spin-orbit interaction on the spin dynamics of a
topologically disordered hopping system is studied in this paper. This is a
significant generalization
of a previous investigation, where an ordered (polaronic) hopping
system has been considered instead.  It is found, that in the limit, where
the Rashba length is large compared to the typical hopping length, the spin
dynamics of a disordered system can still be described by the expressions
derived for an ordered system, under the provision that one takes into
account the frequency dependence of the diffusion constant and the mobility
(which are determined by charge transport and are \emph{independent} of
spin). With these results we are able to make explicit the influence of
disorder on spin related quantities as, e.g., 
the spin life-time in hopping systems.
\end{abstract}

\maketitle

\section{Introduction}

Much attention has recently been devoted to the problem of utilizing electron
spin in semiconductor electronics. 
An overview of this evolving subject of spintronics is
given in Ref.\ \onlinecite{ZuticFabianDasSarma04}.
One central aim of these efforts is the (preferably) electrical
generation, manipulation, and detection of spin currents. 

While a large part of spintronics related literature is concerned with
itinerant electron systems, the corresponding behavior of localized
electrons is worth serious consideration, too.
For instance, some proposals for solid state
quantum computing are based on manipulating the electronic or nuclear spin
state of impurities.\cite{Kane98,OBrienEtal01,
HollenbergEtal04,JelezkoEtal04}

One possibility of affecting the spin behavior through electrical means is
by utilizing two-dimensional structures (semi-conductor interfaces or
heterostructures) showing Rashba spin-orbit interaction
(SOI).\cite{Rashba60}
This paper
considers a model, which consists of a two-dimensional system of localized
electronic states. Charge transport is possible through thermal activation
(hopping transport).\cite{BoettgerBryksin85} The spin dynamics is assumed to
be determined by Rashba SOI. 

In a former
publication\cite{DamkerBoettgerBryksin04} we derived microscopic and
macroscopic transport equations for charge and spin of an ordered
(polaronic) hopping system with Rashba SOI. Here, we generalize 
this theory to
the case of spatial (i.e.\ not energetic) disorder. A suitable reference
system would be nearest-neighbor hopping between donors in a
semi-conductor or between Anderson localized states. The central
approximation concerning spin behavior in hopping systems, which applies to
this investigation, consists in the assumption, that the spin degree of
freedom is solely influenced by Rashba SOI, i.e., there must be no spin
dependent scattering (only s-wave scattering, no magnetic scatterers).
Furthermore, the Rashba length, i.e.\ the length scale over which the spin
rotation occurs, must be large compared to a typical hopping length $R_t$.

In Sec.\ \ref{sectrans} the (disorder averaged) transport equations for spin
are derived for a disordered hopping system and their relation to the charge
transport equations is studied. 
The (averaged or macroscopic)
evolution equations governing charge transport (drift-diffusion
equations) keep their functional shape in the transition from an ordered to a
disordered system, the only effect being, that the diffusion constant and
the mobility become frequency dependent (or, expressed differently: the
macroscopic
evolution equations for the disordered system have memory, even
if the microscopic equations did not).

We show, that the equations for spin dynamics
due to Rashba SOI
likewise keep their functional shape in the presence of disorder, provided
the hopping length is sufficiently short.
Furthermore, the only differences to the ordered case are again a frequency
dependence of the diffusion constant and the mobility of the \emph{charge}.
Thus, provided the length scale for spin inversion due to Rashba SOI (which
will be denoted as the
``Rashba length'' in the following) is
large in comparison with the typical hopping length, the influence of
disorder on the spin dynamics is
entirely determined by the corresponding change of the 
charge transport coefficients.
 
Two different situations are considered in detail: 
The temporal evolution of an initial spin polarization is
investigated (Sec.\ \ref{secinipol}) and the stationary state of a system
with spin polarized boundary conditions is determined 
(Sec.\ \ref{secboundcond}).
In both cases, the analytical treatment is compared with numerical
simulations. 

It is found, that the spin life-time strongly increases with increasing
disorder.  On the other hand, the spatial behavior of the diffusing spin in
the stationary case is practically unaffected by the introduction of
disorder. Additionally, the finding, previously made for an ordered Rashba
hopping system, that there is a critical electrical field, above which the
total spin polarization has an oscillatory component, is also stable against
the introduction of topological disorder.

Section \ref{secdiss} gives a summary of the results and discusses the
experimental relevance. The two appendices present details of the derivation
of the macroscopic spin evolution equation and its long wave-length limit.

\section{The transport equations}
\label{sectrans}

The equations which govern the evolution of charge and spin density of a
hopping system with Rashba SOI have been derived in Refs.\
\onlinecite{DamkerBoettgerBryksin04,DamkerBryksinBoettger04}.
These microscopic transport equations are rate equations
for the occupation numbers and the spin expectation values
of the localized sites.\cite{BoettgerBryksin85}
It is found, that the transition matrix elements between different localized
sites obtains the spin-dependent factor
$\exp\left(-i
\boldsymbol{\sigma}\cdot\left(\boldsymbol{K}\times\boldsymbol{R_{m'm}}\right)
\right)$ due to Rashba SOI. Here,
$\boldsymbol{\sigma}$ denotes the vector of Pauli spin matrices, ${\bf
R}_{m^\prime m}={\bf R}_{m^\prime}-{\bf R}_m$ is the distance vector 
between sites $m^\prime$ and
$m$, and $\boldsymbol{K}$ is a vector normal to the two-dimensional plane,
the length of which is proportional to the Rashba coupling strength. 
Its inverse $1/K=1/\vert{\bf K}\vert$ has the
dimension of a length and will in the following be called Rashba length.

The derivation of the rate equations is subject to the following
restrictions: (i) single particle approximation (i.e.\ no electron-electron
interaction), (ii) Markovian rate equations (i.e.\ the relevant times are
beyond the quantum kinetic regime), and (iii) the Rashba length is large
compared with typical hopping length scales,
$KR_t\ll 1$.
In this paper, we furthermore restrict ourselves to: (iv) two site hopping
probabilities. This gives the leading contribution to hopping transport,
the small parameter being the ratio between the resonance integral and the
energetic width of the set of electron states. Ohmic charge transport is
adequately described by this approximation, but higher order effects, as
e.g. the Hall effect, are neglected.

The corresponding rate equations read\cite{DamkerBoettgerBryksin04}
\begin{equation}
\frac{d}{dt}\rho_m=\sum_{m_1}\left\{\rho_{m_1}W_{m_1m}-\rho_mW_{mm_1}
\right\}
\label{raterho2m}
\end{equation}
for the occupation numbers and
\begin{equation}
\frac{d}{dt}\boldsymbol{\rho}_m=\sum_{m_1}\left\{
\hat{D}_{m_1m}\cdot\boldsymbol{\rho}_{m_1} W_{m_1m}
-\boldsymbol{\rho}_m W_{mm_1}
\right\}
\label{ratevecrho2m}
\end{equation}
for the spin orientation. Here, $\rho_m$ is the occupation number of site
$m$, $\boldsymbol{\rho}_m$ is the expectation value of the spin operator on
this site, $W_{m_1m}$ is the transition rate between sites $m_1$ and $m$,
and $\hat{D}_{m_1m}$ is a $3\times 3$-matrix, which describes a rotation of 
the spin during the transition. An explicit
expression of this matrix is given in Eq.\ (\ref{dofr}). 
Note, that the two sets
of equations are decoupled in the chosen approximation (due to the
restriction to two site hopping probabilities). This means, that the charge
(or particle) transport has no influence on the spin transport and vice
versa. Thus, such an interesting phenomenon as 
the spin Hall effect,\cite{DyakonovPerel71,Edelstein90,Hirsch99}
which already has been studied for ordered hopping
systems\cite{DamkerBoettgerBryksin04}
and which has sparked controversial discussions 
recently,\cite{InoueBauerMolenkamp03,Rashba03,
SinovaEtal04,MurakamiNagaosaZhang04,SchliemannLoss04}
lies outside 
the scope of the theory presented here.

Reference \onlinecite{DamkerBoettgerBryksin04} utilized these rate equations
in order to study the behavior of an ordered (polaronic) hopping system.
Here, a topologically disordered hopping system is the subject of
investigation. Thus, a disorder average has to be performed, in order to
obtain macroscopic transport equations. 

Appendix \ref{appdprod} gives the technical details of the derivation of the
macroscopic transport equations subject to the restrictions given above.
It proceeds along the lines of a ``generic'' averaging procedure, following
the approach used in Ref.\ \onlinecite{KlafterSilbey80}. 

Since the disorder averaged system is homogeneous, it is convenient to work
in wave-vector space. 
The disorder averaged rate equations in Fourier-Laplace space are obtained
as
\begin{equation}
\left[s+W(s\vert{\bf 0})-W(s\vert{\bf q})\right]\rho(s\vert{\bf q})
=\rho_0({\bf q})
\label{rhosq}
\end{equation}
and
\begin{multline}
\left[s+W(s\vert{\bf 0})-\int d^2{\bf q}_1W(s\vert{\bf q}-{\bf q}_1)
\hat{D}({\bf q}_1)
\right]\cdot\boldsymbol{\rho}(s\vert{\bf q})\\
=\boldsymbol{\rho}_0({\bf q}).
\label{vecrhosq}
\end{multline}
Here, $s$ is the Laplace variable (related to time) and ${\bf q}$ is the
(in-plane, i.e.\ two-dimensional) wave vector.

In the long wave-length limit the transition rates of an isotropic system 
can be expanded in the form
\begin{equation}
W(s\vert{\bf q})=W(s\vert{\bf 0})-D(s){\bf q}^2
-{\rm i}\mu(s){\bf q}\cdot{\bf E},
\label{wsqisotrop}
\end{equation}
where the (generally frequency dependent) in-plane 
diffusion constant $D$ and mobility $\mu$ have been introduced. 
The quantity ${\bf E}$ is the in-plane electric
field. In a system of reduced symmetry, $D$ and $\mu$ are tensors, but this
case is not considered here. Note, that the asymmetry between in-plane and
normal-to-plane diffusion and drift does not show here, since the wave
vectors ${\bf q}$ are restricted to two dimensions. 

As shown in App.\ \ref{apprate} the support of $\hat{D}({\bf q})$ lies
within $q<2K$. Thus, provided $K$ is sufficiently small, the long
wave-length limit is also applicable to the integrand in Eq.\
(\ref{vecrhosq}). The condition $KR_t\ll 1$ already assures this, since
wave-vectors corresponding to the (inverse of the) typical hopping length
are within the long wave-length limit. This can safely be assumed, 
because already
for charge transport, we are not concerned with effects due to higher than
the second spatial derivative of $\rho({\bf r})$ in the corresponding 
evolution equation.

Using expression (\ref{wsqisotrop}), 
the integral in Eq.\ (\ref{vecrhosq}) can be
transformed into the $3\times 3$-matrix
\begin{multline}
\Big[(W(\boldsymbol{0})-Dq^2-i\mu{\bf q}\cdot{\bf E})\hat{D}({\bf r})\\
-(2{\rm i}D{\bf q}-\mu{\bf E})\cdot\boldsymbol{\nabla}\circ\hat{D}({\bf r})
+D\Delta\hat{D}({\bf r})\Big]_{{\bf r}=\boldsymbol{0}},
\label{ratevec1}
\end{multline}
where the $s$-dependence has been dropped from the notation. The symbol
$\circ$ denotes the dyadic product.
Thus, only the first three moments of the rotation matrix
$\hat{D}({\bf q})$ (expressed here as spatial derivatives at the origin of
space) enter the
calculations in the long wave-length limit.

Finally, inserting the rotation matrix $\hat{D}$ 
as given by Eq.\ (\ref{dofr})
one obtains the rate equations in the long wave-length limit (see App.\ 
\ref{apprate} for details of the calculation, $\hat{I}_3$ is the $3\times 3$
identity matrix)
\begin{equation}
\Big[s+D(s)q^2+{\rm i}\mu(s){\bf q}\cdot{\bf E}\Big]\rho(s\vert{\bf q})
=\rho_0({\bf q})
\label{raterhoq}
\end{equation}
and
\begin{multline}
\Big[\left(s+D(s)q^2+{\rm i}\mu(s){\bf q}\cdot{\bf E}
+4D(s)K^2\right)\hat{I}_3\\
+4D(s){\bf K}\circ{\bf K}\Big]\cdot\boldsymbol{\rho}(s\vert{\bf q})\\
+\Big[{\bf K}\times\left(
4{\rm i}D(s){\bf q}-2\mu(s){\bf E}\right)\Big]\times
\boldsymbol{\rho}(s\vert{\bf q})
=\boldsymbol{\rho}_0({\bf q}).
\label{raterhovecq}
\end{multline}

These evolution equations agree exactly with the corresponding
equations of an ordered hopping system, derived in Ref.\
\onlinecite{DamkerBoettgerBryksin04}, except that here the diffusion
constant $D$ and the mobility $\mu$ are frequency dependent.  This frequency
dependence is entirely determined by charge transport and does not depend on
SOI within the approximations considered here. 
Thus, one finds that under the
condition $KR_t\ll 1$, the spin dynamics is affected by disorder only
through the change of $D$ and $\mu$. Note, that the above derivation is
independent of the approximations used to derive $W(s\vert{\bf q})$ (the
procedure used to obtain concrete expressions for $D(s)$ and
$\mu(s)$). 
In particular, it is \emph{not} restricted to continuous time random walk,
the procedure further considered in Ref.\ \onlinecite{KlafterSilbey80}, even
though we followed the general approach used there.

For reference, the transport equations (\ref{raterhoq}) and 
(\ref{raterhovecq}) transformed to real space- and time-coordinates read
\begin{widetext}
\begin{equation}
\frac{d}{dt}\rho(t\vert{\bf r})
=\int^t dt_1\Big[D(t-t_1)\Delta\rho(t_1\vert{\bf r})
-\mu(t-t_1){\bf E}\cdot\boldsymbol{\nabla}\rho(t_1\vert{\bf r})\Big],
\label{rhooftint}
\end{equation}
\begin{multline}
\frac{d}{dt}\boldsymbol{\rho}(t\vert{\bf r})
=\int^t dt_1\Big\{
D(t-t_1)\Delta\boldsymbol{\rho}(t_1\vert{\bf r})
-\mu(t-t_1){\bf E}\cdot\boldsymbol{\nabla}\boldsymbol{\rho}(t_1\vert{\bf r})\\
-4K^2D(t-t_1)\boldsymbol{\rho}(t_1\vert{\bf r})
-4K^2D(t-t_1){\bf e}_z\rho_z(t_1\vert{\bf r})\\
-4D(t-t_1)K\left[\boldsymbol{\nabla}\rho_z(t_1\vert{\bf r})-{\bf
e}_z\boldsymbol{\nabla}\cdot\boldsymbol{\rho}(t_1\vert{\bf r})\right]
-2\mu(t-t_1)K\left[{\bf e}_z{\bf E}\cdot\boldsymbol{\rho}(t_1\vert{\bf r})
-{\bf E}\rho_z(t_1\vert{\bf r})\right]
\Big\}.
\label{rhovecoftint}
\end{multline}
\end{widetext}

\section{Total Spin Evolution}
\label{secinipol}
In this section we consider the temporal evolution of the total spin
magnetization. In this case we have to set ${\bf q}={\bf 0}$ in Eqs.\
(\ref{raterhoq}) and (\ref{raterhovecq}). Then, Eq.\ (\ref{raterhoq})
immediately yields particle number conservation. Writing the corresponding
equation for spin polarization in matrix form, while fixing the coordinate
system such that ${\bf E}\parallel{\bf e}_x$ and ${\bf K}\parallel{\bf
e}_z$,
one obtains
\begin{equation}
\begin{bmatrix}
s+4DK^2&0&-2\mu KE\\
0&s+4DK^2&0\\
2\mu KE&0&s+8DK^2
\end{bmatrix}
\cdot\boldsymbol{\rho}(s)=\boldsymbol{\rho}_0.
\label{rhovectotal}
\end{equation}
One can see, that, in zero electric field $E=0$, the spin components decay
with two different time constants (taking $D$ for the moment as frequency
independent): The $z$-component with $\tau_1=1/8DK^2$,
and the in-plane components with twice this value $\tau_2=1/4DK^2$, i.e.\ 
the spin component perpendicular to the plane decays two times faster than
the in-plane components, a fact which has also been found for ordered
hopping systems\cite{DamkerBoettgerBryksin04} and itinerant
electrons.\cite{Bleibaum04,BurkovNunezMacDonald04} 
Note, in particular, that the spin life-time is inversely proportional to
the diffusion constant. Since $D$ substantially decreases with increasing
disorder in hopping systems, the spin life-time will strongly increase with
increasing disorder. 

The frequency dependence of $D$ complicates the matter,
but its overall effect is to further increase the decay time constant for
later times (see the discussion in Sec.\ \ref{secnumerics}). 
Thus, the spin decay slows down further in the progress of time.

Even in a finite electric field, 
the in-plane component of $\boldsymbol{\rho}$ perpendicular to the field
follows a decay law [$\rho_y$ in Eq.\ (\ref{rhovectotal})], 
whereas the other two components are coupled and have the solution
\begin{equation}
\begin{pmatrix}
\rho_x(s)\\
\rho_z(s)
\end{pmatrix}
=
\frac{1}{\det(\Pi)}\Pi
\cdot
\begin{pmatrix}
\rho_{x0}\\
\rho_{z0}
\end{pmatrix},
\label{rhoxztotal}
\end{equation}
with the matrix
\begin{equation}
\Pi=
\begin{bmatrix}
s+8DK^2&2\mu EK\\
-2\mu EK&s+4DK^2
\end{bmatrix}.
\end{equation}

When $D$ and $\mu$ do not depend on frequency, this corresponds to a sum of
exponential functions in time, so that the spin components are either
hyperbolic or trigonometric functions of time, times an exponential decay
factor (see Ref.\ \onlinecite{DamkerBoettgerBryksin04}). Here, in the
disordered case, $D$ and $\mu$ are frequency dependent, and the
transformation of the solution of Eq.\ (\ref{rhovectotal}) into
$t$-space is impossible without choosing a specific model for the frequency
dependence $D(s)$ and $\mu(s)$ beforehand.

For large times, $D$ and $\mu$ approach their respective DC-values, i.e.,
the asymptotic behavior of $\boldsymbol{\rho}$ for large times is easily
obtained. On the other hand, due to the exponential decay, the large time
behavior is only relevant, if the time constant for DC-behavior is smaller
than the time constant for spin decay $1/4DK^2$.

The solution for initial condition $\boldsymbol{\rho}_0={\bf e}_z$ with
time-independent $D$ and $\mu$
is given for future reference
\begin{eqnarray}
\rho_x(t)&=&e^{-6DK^2t}\frac{\epsilon}{\sqrt{\epsilon^2-1}}
\sin(2DK^2\sqrt{\epsilon^2-1}t)
\label{rhoxt}\\
\rho_z(t)&=&e^{-6DK^2t}\Big[\cos(2DK^2\sqrt{\epsilon^2-1}t)\nonumber\\
&&-\frac{1}{\sqrt{\epsilon^2-1}}\sin(2DK^2\sqrt{\epsilon^2-1}t)
\Big].
\label{rhozt}
\end{eqnarray}
The dimensionless electric field is $\epsilon=\mu E/(DK)$. This quantity is
time-independent if the Einstein relation between $D$ and $\mu$ is valid,
even when $D$ and $\mu$ themselves depend on time. In this case, $\epsilon$
furthermore does not depend on the disorder, which also strongly affects the
quantities $D$ and $\mu$ themselves.

The solution here is written
with trigonometric functions, which is the appropriate form for $\epsilon>1$.
In the case $\epsilon<1$, hyperbolic functions have to be used instead.
Thus, the dimensionless electric field $\epsilon$ discriminates between two
different behaviors of the total spin polarization: exponential decay in 
small electric fields and an additional oscillation in large electric
fields. The occurrence of one or the other regime can be tuned by varying
(one or several of) the electric field, the temperature, and the Rashba
length. 

Note, that the conclusions of this section remain the same, whether one
considers a single polarized spin initially placed at the origin, or a
homogeneously polarized system.

\section{Stationary State with Boundary Conditions}
\label{secboundcond}

As a second case, we determine the stationary state of a system with an
in-plane electric field and spin injection. Taking, as before,
${\bf E}\propto{\bf e}_x$, and 
boundaries parallel to the $y$-axis, the charge and spin densities can only
depend on the $x$-coordinate. Denoting the derivative with respect to $x$ by
a prime, one obtains
\begin{equation}
0
=D_0\rho^{\prime\prime}(x)
-\mu_0E\rho^\prime(x)
\label{rhostatic}
\end{equation}
\begin{multline}
0
=D_0\boldsymbol{\rho}^{\prime\prime}(x)
-\mu_0E\boldsymbol{\rho}^\prime(x)
-4D_0K^2\boldsymbol{\rho}(x)
-4D_0K^2{\bf e}_z\rho_z(x)\\
-4D_0K\left[{\bf e}_x\rho_z^\prime(x)-{\bf
e}_z\rho_x^\prime(x)\right]\\
-2\mu_0KE\left[{\bf e}_z\rho_x(x)
-{\bf e}_x\rho_z(x)\right].
\label{vecrhostatic}
\end{multline}
Here, $D_0=\int_0^\infty dtD(t)$ and $\mu_0=\int_0^\infty dt\mu(t)$ are the
DC-values of the corresponding quantities. 

Specifically, if one considers a half-plane, 
and takes the boundary conditions at $x=0$ and $x=\infty$
to be 
$\rho(0)=\rho(\infty)=\rho_0$, $\boldsymbol{\rho}(0)={\bf e}_z$, and
$\boldsymbol{\rho}(\infty)=0$,
the solution then reads $\rho(x)=\rho_0$, 
\begin{eqnarray}
\rho_x(x)&=&e^{-x/\lambda}A_x
\sin(x/\Lambda),
\nonumber\\ 
\rho_y(x)&=&0,
\nonumber\\
\rho_z(x)&=&e^{-x/\lambda}\left[\cos(x/\Lambda)
-A_z
\sin(x/\Lambda)
\right],
\label{rhovecofx}
\end{eqnarray}
where the dimensionless electric field $\epsilon=\mu_0E/(D_0K)$ assumes the
DC-value and is the parameter determining the quantities
$\omega_\pm=\sqrt{\pm\frac{\epsilon^2-8}{2}
+\frac{1}{2}\sqrt{\epsilon^4+48\epsilon^2+512}}$, the amplitudes 
$A_x=\frac{\epsilon^2+32}{5\omega_-+\sqrt{\epsilon^2+7}\omega_+}$ and
$A_z=\frac{5\omega_+-\sqrt{\epsilon^2+7}\omega_-}
{5\omega_-+\sqrt{\epsilon^2+7}\omega_+}$, the decay length 
$\lambda=2(\omega_+-\epsilon)^{-1}/K$ and the oscillation length
$\Lambda=2/(\omega_-K)$.
Note, that the disorder enters only through the ratio $\mu_0/D_0$. 
Thus, provided 
the Einstein relation between $D_0$ and $\mu_0$ is valid, the spatial
behavior of $\boldsymbol{\rho}(x)$ does not depend on the disorder, 
the relevant length scale being determined
mainly by $K$, since the (dimensionless) quantity $\omega_-$ only very weakly
depends on its sole parameter $\epsilon$ (specifically, $3.95<\omega_-\le 4$).

\section{Numerical Simulation}
\label{secnumerics}

Two kinds of numerical simulations are performed. First, the temporal
evolution of a single spin is determined (calculation A, corresponding to
Sec.\ \ref{secinipol}) by solving the
differential equations. This yields, after ensemble averaging, the
temporal behavior of an initially localized $z$-spin. Secondly, the
system is subjected to boundary conditions, which correspond to spin
injection, and the spatial behavior of spin polarization in the stationary
state is determined (calculation B, corresponding to Sec.\
\ref{secboundcond}).  This is done by solving a linear
system of equations. Again, a disorder average must then be performed.  The
disorder averages are performed over about 50 or 1000 realizations for
calculation A or B, respectively. In each case, the statistical error is
of the order of the symbol size in the figures. The number of sites in the
calculations varies
between about 1600 and 5000.  It has been taken care, that the results are
not affected by boundary effects.

The transition rates for the simulation are taken to be 
\begin{equation}
W_{m_1m}=\nu_0e^{-\alpha R_{m_1m}},
\end{equation}
where $\nu_0$ is the attempt frequency (subsuming, e.g., the temperature
dependence of the rates) and determines the numerical time scale. The
parameter $\alpha$ is the inverse localization length of a single impurity
state.
Denoting by $\mathcal{N}$ the (areal) density of localized states, the
quantity $\alpha\mathcal{N}^{-1/2}$ provides a measure of the disorder.

Calculations A and B complement each other. 
Calculation A allows to study the temporal
behavior, but the numerical complexity forbids to investigate systems with
large disorder (large $\alpha\mathcal{N}^{-1/2}$). Large disorder implies a
large dispersion (typically many orders of magnitude) of relevant 
hopping times. Since the numerical time step for a differential equation
solver has to be related to the smallest relevant time scale, but the
long time physics is related to the largest relevant time scale, the study
of systems with large disorder leads to an enormous increase in the number
of time steps to be calculated, if one is interested in the long time
behavior. Furthermore, for larger disorder, the typical behavior can only be
found in systems, which are spatially more extended. This increases the
numerical expense even further. Thus, only relatively small values of
$\alpha\mathcal{N}^{-1/2}$ and/or $t\nu_0$ 
can be investigated by a numerical integration of the rate equations.

On the other hand, calculation B immediately gives the $t\to\infty$ behavior
with a reasonable effort even for systems with large disorder. However, the
temporal behavior is not accessible.

\begin{figure}[!htb]
\includegraphics{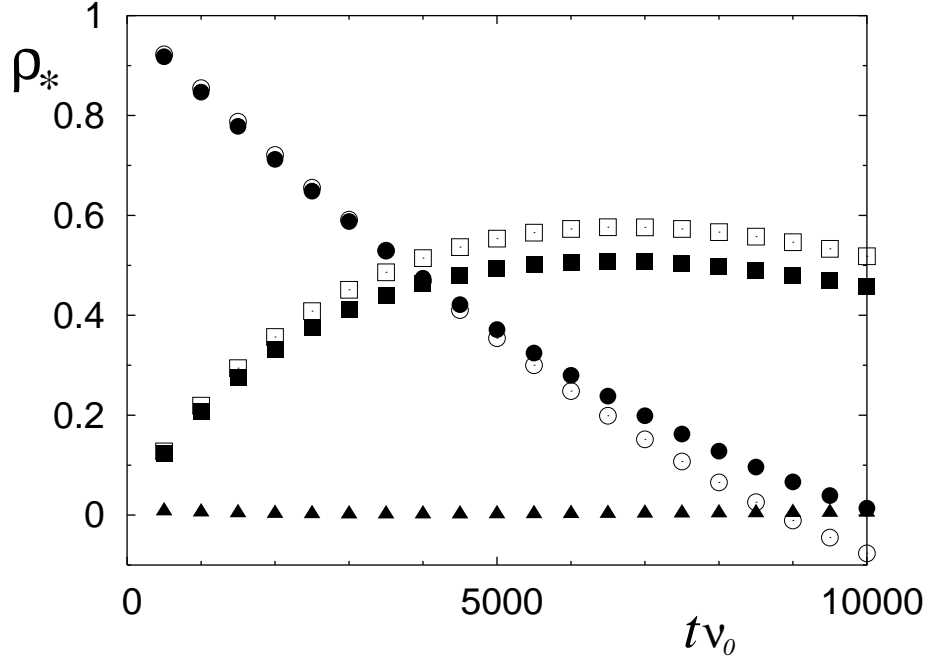}
\caption{Temporal evolution of the components of the total spin polarization 
for a topologically disordered hopping system from numerical simulation:
$\rho_x(t)$ (filled squares), $\rho_y(t)$ (filled triangles), 
$\rho_z(t)$ (filled circles).
For comparison the values computed from the analytical expressions
(\ref{rhoxt}) and (\ref{rhozt}), where $D$ and $\mu$ are replaced by the 
corresponding 
time-dependent quantities $D_{\rm eff}(t)$ and $\mu_{\rm eff}(t)$
[obtained numerically from the charge transport in the same ensemble of
disordered
systems], are also displayed: $\rho_x$ (empty squares), $\rho_y\equiv 0$, 
$\rho_z$ (empty circles).
The statistical error is of the order of the symbol size. 
The parameters are $\alpha\mathcal{N}^{-1/2}=3$, $\epsilon=20$, and 
$K=\sqrt{\mathcal{N}}/10$.}
\label{fig1}
\end{figure}

\begin{figure}[!htb]
\includegraphics{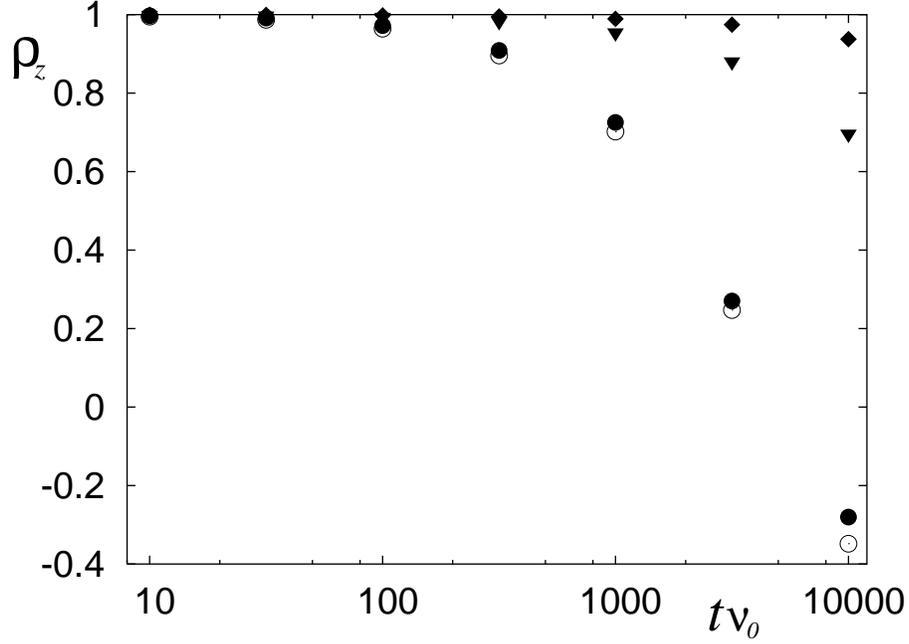}
\caption{Temporal evolution of the $z$-component of the total spin
polarization for
three values of $\alpha\mathcal{N}^{-1/2}=$ 
3 (full circles), 4 (full triangles), 5 (full diamonds). 
For comparison, the analytic solution
for $\alpha\mathcal{N}^{-1/2}=3$ is also given (empty circles).
A quite high value 
of the electric field 
($\epsilon=200$) has been chosen in order to numerically reach the
oscillatory regime (Note the negative value of $\rho_z$ at large times,
which is not possible for plain decay.).
In contrast to Fig.\ \ref{fig1}, the time axis is logarithmic.
Thus, with increasing disorder the spin life-time is strongly enhanced.
$K=\sqrt{\mathcal{N}}/10$.} 
\label{fig2}
\end{figure}

\begin{figure}[!htb]
\includegraphics{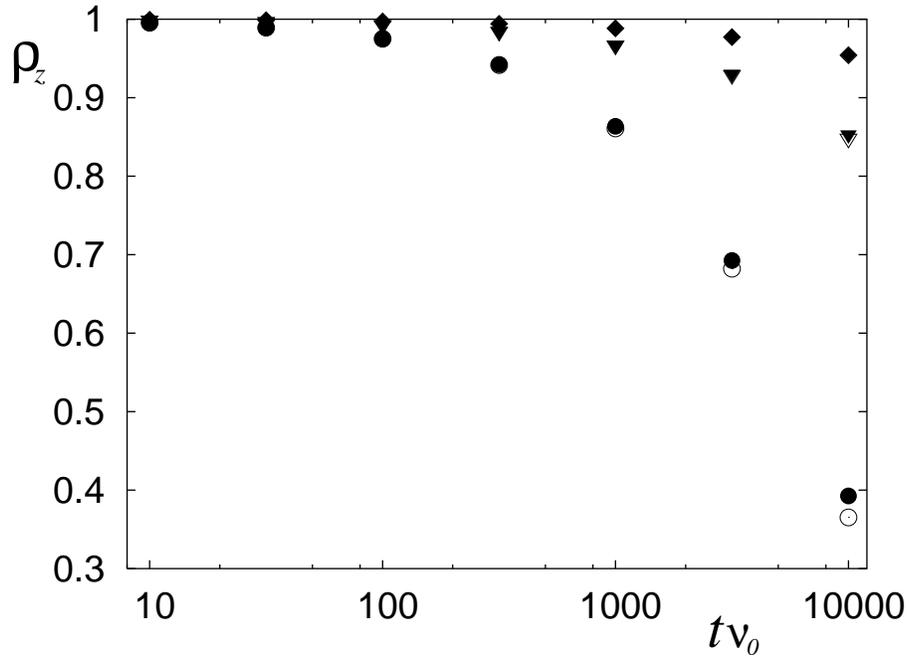}
\caption{Temporal evolution of the $z$-component of total spin 
polarization for $\epsilon=0$. Numerical simulation (full symbols) for
$\alpha\mathcal{N}^{-1/2}=$ 3 (circles), 4 (triangles), 5 (diamonds) and
analytical calculations (empty symbols, respectively) are
shown.
(On this scale, the empty diamonds are indistinguishable from the full ones.)
$K=\sqrt{\mathcal{N}}/10$.}
\label{fig2a}
\end{figure}

To be specific, in calculation A we take the initial condition 
$\boldsymbol{\rho}_0={\bf e}_z$. 
That means, we consider a system with an initial spin polarization in the
$z$-direction. The three components of the 
calculated total spin polarization are shown in Fig.\ \ref{fig1} for the
disorder parameter $\alpha\mathcal{N}^{-1/2}=3$. Figures \ref{fig2} and
\ref{fig2a} give the
$z$-component for several disorder parameters. 

For a comparison with the analytical expressions, the time-dependent
diffusivity and mobility have to be determined. In method A, where the time
evolution of each system is known, effective quantities are determined by
$D_{\rm eff}(t)=\langle{\bf r}^2\rangle/(4t)
=\langle\sum_m\rho_m(t){\bf R}_m^2\rangle/(4t)$ and 
$\mu_{\rm eff}(t)=\langle x\rangle/(Et)
=\langle\sum_m\rho_m(t){\bf R}_m\cdot{\bf e}_x\rangle/(Et)$
from charge diffusion and drift. The brackets $\langle\cdots\rangle$
denote the disorder average.
Method B poses a problem here, since the diffusion co-efficient cannot be
directly measured in this case.
Fortunately only the quotient $\mu/D$, which could be obtained using the 
Einstein relation, enters the analytic expressions.

The relation between the effective quantity $D_{\rm eff}(t)$
and those occurring in the analytical expressions [$D(t)$, the inverse Laplace
transform of $D(s)$] is
unfortunately not simple, but rather involves differentiation operations:
$D(t)=\frac{d^2}{dt^2}[tD_{\rm eff}(t)]$. A corresponding expression connects
$\mu_{\rm eff}$ and $\mu$. Numerical differentiation of noisy data 
(the disorder
averaged $D_{\rm eff}$) is highly unstable, so that the direct comparison of
the numerical results and the analytic expressions (but using the 
numerically obtained $D$ and $\mu$) is not reasonable. 
The situation were much improved, if a
general model for the behavior of $D(t)$ for the system under consideration
were available, but no such model is available for the range of parameters
considered here. 

\begin{figure}[!htb]
\includegraphics{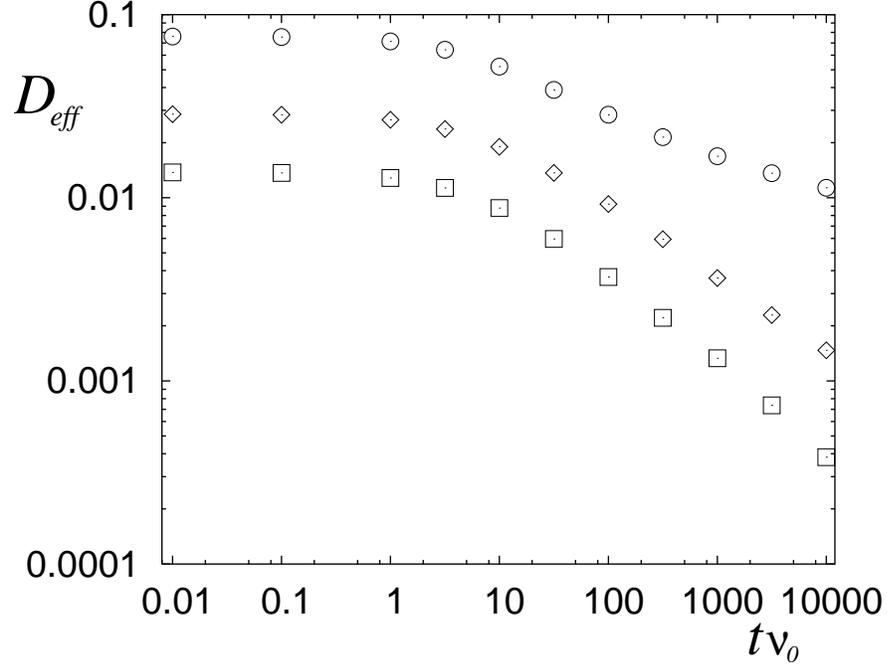}
\caption{Time dependence of the effective diffusion constant 
$D_{\rm eff}(t)$ for disorder parameters
$\alpha\mathcal{N}^{-1/2}=$ 3 (circle), 5 (diamond), and 7 (square),
numerically determined from charge transport in a topologically disordered
hopping system. The scale of the ordinate is arbitrary.}
\label{fig1a}
\end{figure}

Thus, we take a different approach. The asymptotic values of
$D_{\rm eff}(t)$ and $\int_0^tdt_1D(t_1)$
are the same for very small ($t\nu_0<1$) or for very large times
(DC-behavior).
In between, for moderate times, the detailed behavior differs, but, since
they change monotonically over several orders of
magnitude\cite{BoettgerBryksin85} (see Fig.\
\ref{fig1a}), we expect the Markovian limit of the transport equations to be
valid in the following sense. 
Because $D(t)$ and $\mu(t)$ decrease over many orders of magnitude with
increasing t,
the main contribution to the time integrals in Eqs.\ (\ref{rhooftint}) and
(\ref{rhovecoftint}) comes from small times $t<1/\nu_0$. 
The charge and spin density
is taken out of the integral (as in the derivation of a Markovian equation),
the remaining time integrals replaced by 
$\int_0^t dtD(t)\approx D_{\rm eff}(t)$, and for $\mu$ accordingly.
The resulting equations are solved analytically, but using the numerically
obtained (from charge transport) quantities $D_{\rm eff}(t)$ and $\mu_{\rm
eff}(t)$.
The analytic solutions calculated in this way are displayed 
in Figs.\ \ref{fig1} to \ref{fig2a}
for comparison. One can see, that the agreement with the
numerical simulation is reasonable.

This discussion can be somewhat improved, without the need to introduce
a specific model for $D(t)$. Figure \ref{fig1a} shows, that approximately
\begin{equation}
D_{\rm eff}(t)=
\begin{cases}
D_\infty,&(\nu_0t<1);\cr
D_\infty(\nu_0 t)^{-\kappa},&(\nu_0t>1);\cr
D_0,&(\nu_0t\gg 1),\cr
\end{cases}
\label{deffapprox}
\end{equation}
with constants $D_\infty$ (infinite frequency) and $D_0$ (DC value).
(The third case cannot really be deduced from Fig.\ \ref{fig1a}, but it is
justified for any system which does not have an infinite memory.)
The exponent is about $\kappa\approx 1/4$ for $\alpha\mathcal{N}^{-1/2}=3$ 
and $\kappa\approx 1/2$ for $\alpha\mathcal{N}^{-1/2}=7$.
Small and large times [the first and third case of Eq.\ (\ref{deffapprox})]
thus permit the approximation $\int_0^tdt_1D(t_1)\approx D_{\rm eff}(t)$.
On the other hand, for medium times [the second case of Eq.\
(\ref{deffapprox}), which corresponds to the relevant time scale of the
numerical calculations], one obtains
\begin{equation}
\int_0^tdt_1D(t_1)\approx(1-\kappa)D_{\rm eff}(t).
\end{equation}
Thus, the effective $D$ for the analytic expression has to be slightly
reduced (by the factor $1-\kappa$). 
This has the effect [see Eqs.\ (\ref{rhoxt}) and (\ref{rhozt})]
of decreasing
the decay rate and also the frequency, in effect moving the zero crossings 
to larger times. Calculations show, that this model of $D_{\rm eff}(t)$ is 
too crude to give a better agreement with the numerics compared with the
simple model advertised in the previous paragraph. The general trend of
diminishing the decay rate and the frequency,
however, is in accord with the numerical evidence (see Fig.\ \ref{fig1}).

Note, that the oscillatory behavior of the total spin polarization for
$\epsilon>1$, which has been predicted for ordered hopping
systems\cite{DamkerBoettgerBryksin04} survives the introduction of disorder.
Furthermore, Figs.\ \ref{fig2} and \ref{fig2a} show, that increasing disorder (increasing
$\alpha\mathcal{N}^{-1/2}$) leads to an increase of the spin life-time. This
can be readily explained by noting the strong decrease of the diffusion 
constant in the transport equations, a conclusion, which is confirmed by the
agreement between numerical simulation and analytical calculation.

\begin{figure}[!htb]
\includegraphics{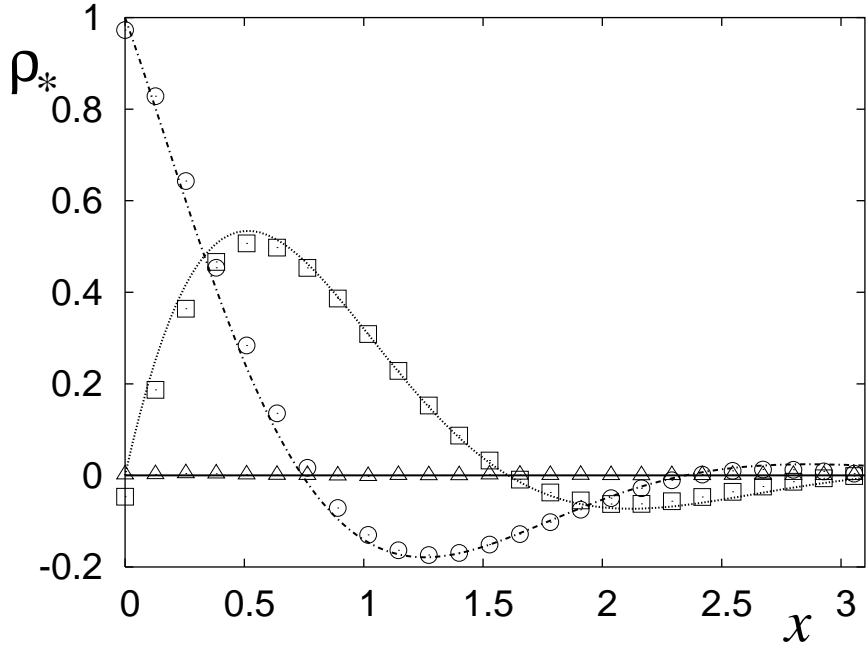}
\caption{Spin components in the stationary state for a 
disordered hopping system with spin injection simulated 
by the boundary condition $\boldsymbol{\rho}(x=0)={\bf e}_z$. 
The results of numerical simulation
$\rho_x$ (square),
$\rho_y$ (triangle), $\rho_z$ (circle),
compare well with the analytical results Eq.\ (\ref{rhovecofx}):
$\rho_x$ (dotted), $\rho_z$ (dash-dotted).
The $y$-components of the spin are zero within the statistical error.
The parameters $\alpha\mathcal{N}^{-1/2}=7$, $\epsilon=0.232$, and 
$K=\sqrt{\mathcal{N}}/20$. The dimensionless length $x=\vert{\bf r}\cdot{\bf
e}_x\vert/K$ measures the
distance from the boundary and the in-plane electrical field is aligned
perpendicular to the boundary at $x=0$.
}
\label{fig3}
\end{figure}

The results of calculation B (spin injection with boundary conditions) are
shown in Fig.\ \ref{fig3}. The only parameter of the analytical expressions
is the ratio $D_0/\mu_0$ which is assumed to take the value given by the
Einstein relation ($kT/e$). The agreement to the analytical expressions
(\ref{rhovecofx}) shown in Fig.\ \ref{fig3} is convincing. Note, that in
this case, no parameter had to be obtained from charge transport. Instead,
the temperature used in the numerical simulation is used in the analytical
calculation, too. Figure \ref{fig3} shows only the results for the single
disorder parameter $\alpha\mathcal{N}^{-1/2}=7$, but the good agreement
between numerics and Eq.\ (\ref{rhovecofx}) holds also for the other
investigated parameters (disorder
$3\le\alpha\mathcal{N}^{-1/2}\le 7$ in small or medium fields
$0\le\epsilon\le1$).

\begin{figure}[!htb]
\includegraphics{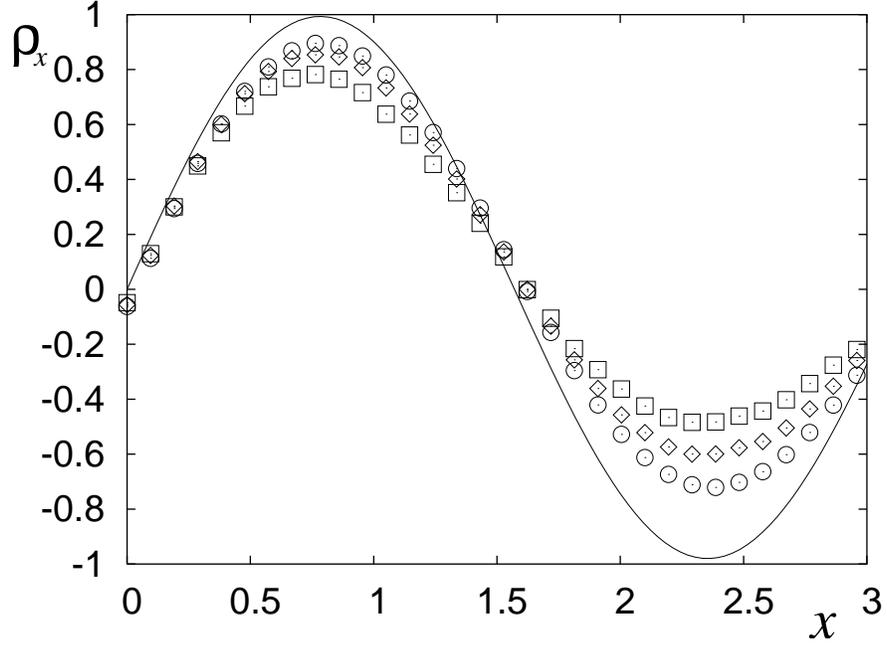}
\caption{The x-component of the spin in the stationary state 
for three values of 
$\alpha\mathcal{N}^{-1/2}=$ 3 (circle), 5 (diamond), 7 (square),
and the analytical result Eq.\ (\ref{rhovecofx}) (full line). 
The situation is analogous to Fig.\ \ref{fig3}, except for a stronger
electric field $\epsilon=232$.
One can see, that for large
$\epsilon$, the magnitude of the spin decreases with increasing disorder,
and differs increasingly from the analytical results. 
(The $z$-component behaves analogously.)
A discussion of this behavior is given in the text.}
\label{fig4}
\end{figure}

A deviation is only found in very large electrical fields ($\epsilon\gg 1$,
see Fig.\ \ref{fig4}).
The spatial spin coherence in calculation B diminishes with increasing
$\alpha\mathcal{N}^{-1/2}$
in large electrical fields. 
The numerical data can still be fitted to the analytical model, but with a
larger value of $\epsilon$ as given by the Einstein relation (not shown).
This indicates, that the ratio $D_0/\mu_0$
increases beyond the equilibrium value $kT/e$ in large fields, but
otherwise, the spin behavior is still described by Eq.\ (\ref{rhovecofx}). 
This is
indicative of a transition from isotropic to directed percolation in a
growing electric field. This transition leaves the diffusion constant 
nearly unaffected, while the mobility decreases. This scenario can explain the
observed behavior qualitatively, but further research is necessary
to illuminate the true physical basis of the observed behavior
and exclude other possible explanations.

\section{Discussion}
\label{secdiss}

We have derived macroscopic spin transport equations for spatially
disordered hopping systems with Rashba SOI. It is found, that the
introduction of disorder leaves the vectorial structure of these equations
intact, the only effect being that diffusion constant and mobility become
frequency dependent (or, expressed differently, that the transport equations
obtain memory). This frequency dependence is already determined by the
charge transport behavior and does not depend on the specifics of spin
transport. The derivation of
this relation between ordered and disordered hopping spin transport is
subject to the approximation, that the Rashba length (the length scale of
spin precession) is large against a typical hopping length. Furthermore,
only two-site hopping processes can be dealt with in the present treatment,
thus excluding, e.g., the discussion of a possible spin Hall effect. 

Other
effects, previously predicted for ordered spin hopping, also occur for
disordered spin hopping. For instance, the spin decay is exponential in
small electric fields, whereas it obtains an oscillatory component in large
electric fields. Thus, there is a finite critical field, dividing both
regimes, also in the disordered case. It is also conceivable that the
behavior ``oscillates'' between exponential and vibrational behavior. But
this can be excluded, so far as $D(t)$ and $\mu(t)$ change monotonically 
with time.

In comparison to the ordered case, the spin life-time is shown to be
strongly increased by the introduction of disorder. 
This is explained by the significantly reduced
diffusion constant, which enters the life-time reciprocally and 
which sharply decreases with increasing disorder.

For the stationary state of spins injected through a boundary into a
topologically disorder hopping system, it is found, that the spatial
behavior of the spin polarization is largely unaffected by varying the
disorder. An exception to this insensitivity is the systematic reduction of
spin polarization with increasing disorder in large in-plane electric fields
$\epsilon\gg 1$. But even here, the numerical data indicate, 
that the phenomenological
description of spin transport, which is derived in this paper, 
is still valid. The deviation appears to be due to the invalidity of the
Einstein relation between diffusion co-efficient and mobility in strong
electric fields for the considered system, and is not a consequence of
the peculiarities of spin transport.

A comment concerning the required smallness of the Rashba interaction
is in order here. For the value $K=\sqrt{\mathcal{N}}/10$ 
(the largest value used in the simulations)
the spin reverses over a distance of
the order of about ten times the mean distance between impurities. The
typical hopping length also is several mean distances long. Thus, the
central condition $KR_t\ll 1$ is hardly valid.
This (large) value of $K$ has been taken for numerical reasons, 
but even so, the agreement with theory is very good. This 
shows, that the phenomenological description of spin transport derived above
is quite robust even for (relatively) large values of the Rashba interaction
strength. Smaller (more realistic) values of the Rashba coupling 
$K$ render the approximation all the more valid.

Finally, a few words concerning the experimental relevance of this work. 
Usually given values for the Rashba SOI strength
($10^{-9}$~eV~cm)\cite{ShahbazyanRaikh94,MolenkampSchmidtBauer01}
lead to
a Rashba length of the order of $1/K\approx 100$~\AA, if one assumes that
the effective electron mass is equal to its elementary
mass.\cite{DamkerBoettgerBryksin04} This length scale is in the range of
typical hopping lengths. Thus, depending on the system under consideration,
the theory presented here is a candidate for a description of the
physical behavior. For the value of the Rashba coupling given above and at a
temperature of 10~K, the critical electrical field $\epsilon=1$, which
differentiates between exponential and oscillatory behavior of  the total
spin polarization, corresponds to a field of about $10^3$~V/cm in natural
units.

To the authors knowledge, 
the value of the Rashba coupling strength in hopping systems has not been
investigated up to now. It may well be, that this value is smaller than for
itinerant electrons, in which case the estimate given above for the Rashba
length (100~\AA) must be increased. Then, it is even easier to comply with
the condition $KR_t\ll 1$. A further advantage of this case would be, that
a modification of the Rashba coupling strength by an external gate
electrode\cite{NittaEtal97}
would require smaller applied voltages. 

\appendix
\section{The macroscopic transport equations for spin density}
\label{appdprod}

The aim is to derive macroscopic transport equations starting from the
microscopic equations (\ref{raterho2m}), the master equation for particle 
(respectively charge) transport, and (\ref{ratevecrho2m}), the equation
governing the spin evolution.
In the following, we show
that under the approximation of a large Rashba length
$KR_t\ll 1$, the disorder average of Eq.\ (\ref{raterho2m}) also gives an
averaged equation for spin, where the averaged equations have the same
relation to each other, as is the case for Eqs.\ (\ref{raterho2m}) and
(\ref{ratevecrho2m}). In order to elucidate the correspondence between the
disorder averages of both equations, in the following, we first consider a
``generic'' averaging procedure for charge transport, and then explore the
outcome of the same procedure as applied to spin transport.

The ``generic'' averaging procedure follows the procedure used by Klafter
and Silbey\cite{KlafterSilbey80} which they employed to study the connection
between the continuous time random walk approach and the exact (generalized)
master equation. We assume, that the spatially disordered system can be
adequately represented using an ordered host lattice with very small lattice
constant. Then only some sites of the host lattice
correspond to the original sites and are available for transport, whereas
all other host lattice sites are unavailable in a specific disorder
realization. The transition rates
$W_{m_1m}$ (where $m$ and $m_1$ correspond to sites of the original
[disordered] model) can be generalized to transition rates $V_{n_1n}$ on the
host lattice, such that
$V_{n_1n}$ vanishes, except when both sites are available as defined above.
Using this ordered representation of the originally disordered problem, one
can construct a formal solution and thereafter perform the disorder average
(see Ref.\ \onlinecite{KlafterSilbey80}). In this way, one obtains a
generalized master equation (GME) for the (averaged or macroscopic)
disordered system, which is formally exact. Approximations (as, e.g., the
continuous time random walk considered in Ref.\
\onlinecite{KlafterSilbey80}) are only needed thereafter, in order to obtain
explicit expressions for the transition rates of the GME and thus be able to
calculate solutions of the GME. In the following paragraph we will show
that, provided $KR_t\ll 1$, the derivation of the formally exact GME for
charge transport [starting from Eq.\ (\ref{raterho2m})] can nearly unaltered
be applied to spin transport [starting from Eq.\ (\ref{ratevecrho2m})], too.
In this way, any approximations, which are applied to the charge transport
GME, yield immediately a corresponding spin transport GME without the need
to introduce further approximations (except $KR_t\ll 1$ of course).

We denote the (disordered) transition rates on the ordered host lattice 
(with site indices $n$, etc.) by
\begin{equation}
V_{n_1n}=W_{m_1m}\delta_{n_1m_1}\delta_{nm}
\end{equation}
and introduce diagonal elements by
\begin{equation}
V_{nn}=-\sum_{m_1}W_{mm_1}\delta_{nm},
\end{equation}
where $\delta_{nm}$ is equal to unity or to zero, 
when the host lattice site $n$
coincides or not with an original site $m$ in the current disorder realization,
respectively.
Then, the charge transport rate equation can be written as
\begin{equation}
\frac{d}{dt}\rho_n(t)=\sum_{n_1}V_{n_1n}\rho_{n_1}.
\end{equation}
These equations can equivalently be expressed as a matrix equation, such
that, if $N$ denotes the number of sites of the host lattice,
the $\rho_n$ become an $N$-vector $\underline{\rho}$ and correspondingly
$V_{n_1n}$ becomes an $N\times N$-matrix $\underline{\underline{V}}$. The
rate equation then takes the simple form
$d\underline{\rho}/dt=\underline{\underline{V}}\,\underline{\rho}$, or, in
Laplace space $[s-\underline{\underline{V}}]\underline{\rho}
=\underline{\rho}_0$, where the index $0$ marks the initial conditions. 
The averaged solution is formally obtained as
\begin{equation}
\langle\underline{\rho}(s)\rangle=
\left\langle[s-\underline{\underline{V}}]^{-1}\right\rangle
\underline{\rho}_0
=[s-\underline{\underline{M}}(s)]^{-1}\underline{\rho}_0,
\end{equation}
where $\langle\ldots\rangle$ denotes disorder average. The second equation
constitutes the definition of the self-energy $\underline{\underline{M}}$,
the elements of which are the (as yet
exact, but only formally defined) transition rates of the GME
\begin{equation}
s\rho_n(s)-\sum_{n_1}M_{n_1n}(s)\rho_{n_1}(s)=\rho_{n0}.
\end{equation}

Conservation of probability allows to write this equation in the usual form
\begin{equation}
s\rho_n(s)-\rho_{n0}
=\sum_{n_1}[M_{n_1n}(s)\rho_{n_1}(s)-M_{nn_1}(s)\rho_n(s)],
\end{equation}
where the sum now excludes the term $n_1=n$. Note, that this equation lives
on an ordered (quasi-continuous) lattice in contrast to Eq.\
(\ref{raterho2m}). One can see, that the formal structure of Eq.\
(\ref{raterho2m}) has survived the averaging, except that the transition
rates have become frequency dependent. Thus, even though the microscopic
equations (\ref{raterho2m}) are Markovian, the macroscopic (averaged)
equations can be non-Markovian.

Let us now execute the same procedure for the spin density. The quantity
$\boldsymbol{\rho}_n$ is a 3-vector, thus $\underline{\boldsymbol{\rho}}$
lives in the Cartesian product space of 3-vectors and $N$-vectors.
The transition rates are products $\hat{D}_{n_1n}V_{n_1n}=\hat{V}_{n_1n}$,
where $\hat{D}_{n_1n}$ determines the action in 3-space and $V_{n_1n}$ is
identical to the corresponding quantity defined above for charge transport.
Note, that we do not need a special rule for the ``diagonal elements'',
because $\hat{D}_{nn}$ is the unity matrix. 
Thus, the formal solution is 
\begin{equation}
\langle\underline{\boldsymbol{\rho}}(s)\rangle=
\left\langle[s-\underline{\underline{\hat{V}}}]^{-1}\right\rangle
\cdot\underline{\boldsymbol{\rho}}_0
=[s-\underline{\underline{\hat{M}}}(s)]^{-1}
\cdot\underline{\boldsymbol{\rho}}_0,
\label{rhovec2}
\end{equation}
defining the quantity $\underline{\underline{\hat{M}}}$.

We now introduce the assumption, that a product of several rotation matrices
$\hat{D}$ obeys the relation
\begin{equation}
\hat{D}_{mm_1}\cdot\hat{D}_{m_1m_2}\cdot\ldots\cdot\hat{D}_{m_kn}\approx
\hat{D}_{mn}.
\label{dproduct}
\end{equation}
This equation is valid to first order in
$K$. Therefore, the condition $KR_t\ll 1$, where $R_t$ is a relevant length
scale of the hopping transport, assures the applicability of the
approximation. The fact, that Eq.\ (\ref{dproduct}) is valid only to first
order in $K$, gives the reason for the restriction to two-site hopping
probabilities in this paper. Three-site probabilities yield interesting
physics\cite{DamkerBoettgerBryksin04} as, e.g., the spin Hall effect,
but those effects are of higher order in $K$.

Within the approximation (\ref{dproduct}), it can be shown, that the spin
dependence of the transition matrix $\underline{\underline{\hat{M}}}$ is
given by 
\begin{equation}
\hat{M}_{n_1n}=M_{n_1n}\hat{D}_{n_1n}
\label{mhatofd}
\end{equation}
and the rate equation of the averaged system reads
\begin{equation}
s\boldsymbol\rho_n(s)-\boldsymbol\rho_{n0}
=\sum_{n_1}[M_{n_1n}(s)\hat{D}_{n_1n}\cdot\boldsymbol\rho_{n_1}(s)
-M_{nn_1}(s)\boldsymbol\rho_n(s)].
\end{equation}

The approximate validity of Eq.\ (\ref{mhatofd})
can be seen by representing the inverse matrix
$[s-(\underline{\underline{\hat{V}}})]^{-1}$ occurring in Eq.\ (\ref{rhovec2})
as a geometric series. In each term of this
series, the product (\ref{dproduct}) occurs, and is accordingly simplified. 
The disorder average then yields an expression with the structure
\begin{equation}
\frac{\hat{D}_{n_1n}}{s}\left(\underline{\underline{1}}_{n_1n}
+\frac{1}{s}\langle\underline{\underline{V}}\rangle_{n_1n}
+\frac{1}{s^2}\langle\underline{\underline{V}}^2\rangle_{n_1n}
+\ldots
\right)
\end{equation}
The series of averaged powers of $\underline{\underline{V}}$ are equal to
the expression
$[s-\underline{\underline{M}}]^{-1}$ from charge transport, 
which can similarly be expanded into a geometric
series, and the overall
$\hat{D}$-factor split [as if Eq.\ (\ref{dproduct}) is read from right to
left], so that each term $\hat{D}_{n_1n}(\underline{\underline{M}}^k)_{n_1n}$ 
of the new geometric series becomes 
$(\underline{\underline{\hat{M}}}^k)_{n_1n}$ and we thereby obtain
Eq.\ (\ref{mhatofd}).

Another way to derive the approximation, which uses the Zwanzig projection
operator method, is the following. 
Introducing a projection
operator $P$ as effecting the disorder average
\begin{equation}
P\underline{\underline{A}}=\langle\underline{\underline{A}}\rangle,
\end{equation}
the self-energy can be expressed as 
(see, e.g., Ref.\ \onlinecite{KlafterSilbey80})
\begin{equation}
\underline{\underline{\hat{M}}}
=\left\langle\underline{\underline{\hat{V}}}\right\rangle
+\left\langle\delta\underline{\underline{\hat{V}}}
\left[s-(1-P)\underline{\underline{\hat{V}}}\right]^{-1}
\delta\underline{\underline{\hat{V}}}\right\rangle,
\end{equation}
where $\delta\underline{\underline{\hat{V}}}=\underline{\underline{\hat{V}}}
-\langle\underline{\underline{\hat{V}}}\rangle$. 
Here, only one term, namely $[s-(1-P)\underline{\underline{\hat{V}}}]^{-1}$,
has to be expanded in a geometric series. The application of
approximation (\ref{dproduct}) again leads to the conclusion (\ref{mhatofd}).

Homogeneity of the averaged system implies, that
$M_{n_1n}$ only depends on the difference vector between the sites
$M_{n_1n}=M({\bf R}_{n_1n})$. Thus, it is convenient to work in wave-vector
space.
The disorder averaged rate equations in Fourier-Laplace space then become
the
Eqs.\ (\ref{rhosq}) and (\ref{vecrhosq}),
where we have replaced the symbol $M$ by $W$, with the understanding that
the quantities $W_{m_1m}$ are the transition rates for the disordered
system, whereas $W(s\vert{\bf q})$ or $W(s\vert{\bf r})$ are the transition 
rates for the averaged --- and thus homogeneous --- system.

\section{The rate equations in the long wave-length limit}
\label{apprate}
The following expressions are written in the right-handed orthonormal basis
${\bf e}_r={\bf r}/r$, ${\bf e}_{K\times r}={\bf K}\times{\bf r}/Kr$, and 
${\bf e}_z={\bf K}/K$.

The spin rotation matrix for a transition between sites $m$
and $m_1$ depends only on the distance
vector between both sites $\hat{D}_{m_1m}=\hat{D}({\bf R}_{m_1m})$. The
explicit expression for this quantity is\cite{DamkerBoettgerBryksin04}
\begin{multline}
\hat{D}({\bf r})=\hat{I}_3
+\sin(2Kr)({\bf e}_r\circ{\bf e}_z-{\bf e}_z\circ{\bf e}_r)\\
{}+(\cos(2Kr)-1)({\bf e}_z\circ{\bf e}_z+{\bf e}_r\circ{\bf e}_r).
\label{dofr}
\end{multline}

First, we show that the Fourier transform $\hat{D}({\bf q})$ has finite
support, specifically, it is restricted to the disc 
$\vert{\bf q}\vert\le 2K$. The (lengthy) calculation is not given here in
detail, but proceeds in the following way: The (divergent) 
Fourier integral of $\hat{D}({\bf r})$ is made convergent by 
introducing the factor $e^{-\alpha r}$ with a positive parameter $\alpha$.
The corresponding transform $\hat{D}_\alpha({\bf q})$ can be calculated
analytically and consists of rational functions of $q$ and $\alpha$, except
for non-analytic factors of $[q^2+\alpha^2]^{-1/2}$ and 
$[q^2+(\alpha\pm 2Ki)^2]^{-1/2}$, which only have cuts and singularities in
the disc
$q\le 2K$. The subsequent limit $\alpha\to 0^+$ shows that $\hat{D}({\bf q})$
is a sum of $\delta$-distributions and their derivatives with support within
this disc.
Specifically, for $q>2K$ one obtains 
$\lim_{\alpha\to 0^+}\hat{D}_\alpha({\bf q})=\hat{0}$,
the zero matrix. 
Thus,
\begin{equation}
\hat{D}({\bf q})=0 \text{ for } q>2K.
\end{equation}

Next, the first and second spatial derivative of
$\hat{D}({\bf r})$ are calculated.
Recall, that the
vector ${\bf r}$ is two-dimensional. Thus, e.g.,
$\boldsymbol{\nabla}\circ{\bf r}={\bf e}_r\circ{\bf e}_r
+{\bf e}_{K\times r}\circ{\bf e}_{K\times r}$, where one should note, that
the $z$-components are missing. Keeping this in mind, we have
$\boldsymbol{\nabla}\circ{\bf e}_r
={\bf e}_{K\times r}\circ{\bf e}_{K\times r}/r$. Of course
$\boldsymbol{\nabla}\circ{\bf e}_z=0$, and thus
$\boldsymbol{\nabla}\circ{\bf e}_{K\times r}
=\boldsymbol{\nabla}\circ({\bf e}_z\times{\bf e}_r)
=-{\bf e}_{K\times r}\circ{\bf e}_r/r$. Replacing the dyadic
product by a scalar product (corresponding to a tensor contraction)
immediately gives the additional relations
$\boldsymbol{\nabla}\cdot{\bf e}_r=1/r$ and $\boldsymbol{\nabla}\cdot{\bf
e}_{K\times r}=0$.

One can now easily calculate
\begin{widetext}
\begin{multline}
\boldsymbol{\nabla}\circ\hat{D}({\bf r})=2K\cos(2Kr){\bf e}_r\circ
({\bf e}_r\circ{\bf e}_z-{\bf e}_z\circ{\bf e}_r)
+\frac{\sin(2Kr)}{r}{\bf e}_{K\times r}\circ({\bf e}_{K\times r}
\circ{\bf e}_z-{\bf e}_z\circ{\bf e}_{K\times r})\\
{}-2K\sin(2Kr){\bf e}_r\circ({\bf e}_z\circ{\bf e}_z+{\bf e}_r\circ{\bf
e}_r)
+\frac{\cos(2Kr)-1}{r}{\bf e}_{K\times r}\circ({\bf e}_{K\times r}
\circ{\bf e}_r+{\bf e}_r\circ{\bf e}_{K\times r})
\end{multline}
and
\begin{multline}
\Delta\hat{D}({\bf r})
=\left(-4K^2\sin(2Kr)+2K\frac{\cos(2Kr)}{r}-\frac{\sin(2Kr)}{r^2}\right)
({\bf e}_r\circ{\bf e}_z-{\bf e}_z\circ{\bf e}_r)\\
-\left(4K^2\cos(2Kr)+2K\frac{\sin(2Kr)}{r}
\right)({\bf e}_z\circ{\bf e}_z+{\bf e}_r\circ{\bf e}_r)
-2\frac{\cos(2Kr)-1}{r^2}({\bf e}_r\circ{\bf e}_r-{\bf e}_{K\times r}\circ
{\bf e}_{K\times r}).
\end{multline}
\end{widetext}

The limit ${\bf r}\to\boldsymbol{0}$ needs a special comment,
because one formally obtains expressions as, e.g.,
\begin{multline}
\boldsymbol{\nabla}\circ\hat{D}({\bf r})\vert_{{\bf r}=\boldsymbol{0}}
=2K{\bf e}_r\circ
({\bf e}_r\circ{\bf e}_z-{\bf e}_z\circ{\bf e}_r)\\
{}+2K{\bf e}_{K\times r}\circ({\bf e}_{K\times r}
\circ{\bf e}_z-{\bf e}_z\circ{\bf e}_{K\times r}),
\end{multline}
which contain the basis vectors ${\bf e}_r$ and ${\bf e}_{K\times r}$,
which are indeterminate in the limit ${\bf r}=\boldsymbol{0}$. This seeming
puzzle
is solved by the observation, that these basis vectors only occur in
combinations, which yield the unit matrix $\hat{I}_2$ in two dimensions. 

Thus, one obtains
\begin{equation}
{\bf a}\cdot\boldsymbol{\nabla}\circ\hat{D}({\bf r})
\vert_{{\bf r}=\boldsymbol{0}}=2K({\bf a}\circ{\bf e}_z-{\bf e}_z\circ{\bf
a}),
\end{equation}
where $\bf a$ is an arbitrary vector, and
\begin{equation}
\Delta\hat{D}({\bf r})\vert_{{\bf r}=\boldsymbol{0}}
=-4K^2\hat{I}_3-4K^2{\bf e}_z\circ{\bf e}_z.
\end{equation}
Inserting these expressions into Eq.\ (\ref{ratevec1}) gives Eq.\
(\ref{raterhovecq}).

\newpage
\onecolumngrid

\end{document}